\documentclass[prl,floatfix,showpacs,twocolumn]{revtex4}
\usepackage{amsmath}
\usepackage{amssymb}
\usepackage{graphicx}

\begin{document}

\title{Entropically driven transition to a liquid-crystalline polymer globule}

\author{C. Nowak}
\author{V. G. Rostiashvili}
\author{T.A. Vilgis}

\affiliation{Max-Planck-Institut f\"ur Polymerforschung, 
             Ackermannweg 10, 
             55128 Mainz, 
             Germany}

%\date{\today}

\begin{abstract}
A self-consistent-field theory (SCFT) in the grand canonical ensemble formulation is used to study transitions in a helix-coil multiblock 
copolymer globule. 
The helices are modeled as stiff rods. In addition to the established coil-globule transition we show for the first time that, even without explicit 
rod-rod  alignment interaction, the system undergoes a transition 
to a nematic liquid-crystalline (LC) globular state. The LC-globule formation 
is driven by  the hydrophobic helical segment attraction and  the anisotropy of the globule surface energy. 
The full phase diagram of the copolymer was calculated. It discriminates between an open chain, 
amorphous globule and LC-globule. 
This model provides a relatively simple example of the interplay between secondary and tertiary structures 
in homopolypeptides. Moreover, it gives a simple explanation for the formation of helix bundles in certain globular proteins. 
\end{abstract}


\maketitle

%%%%%%%%%%%%%%%%%%%%%%%%%%%%%%%%%%%%%%%%%%%%%%%%%%%%%%%%%%%%%%%%%%%%%%%%%%%
%%%%%%%%%%%%%%%%%%%%%%%%%%%%%%%%%%%%%%%%%%%%%%%%%%%%%%%%%%%%%%%%%%%%%%%%%%%
The formation of secondary structure in proteins and in particular the $ \alpha $-helix-coil transition in homopolypeptide chains is one of the well-investigated conformational transitions in biomolecules. In the well-known Zimm-Bragg (ZB)-theory of this phenomenon \cite{Zimm,Poland} the polypeptide molecule is considered as a one-dimensional cooperative system and the problem can be solved exactly. However, ZB-theory is only valid in a fully denaturated state when all three dimensional interactions are negligible. In real systems the helical parts are often hydrophobic, such that this hydrophobicity drives the helix-coil copolymer into a globular phase.
Furthermore, helical parts can be seen as rigid rods and hence a rod-rod  alignment interaction could be taken into account. Considering all these facts, it is natural to pose the question: how does the chain compaction affect the secondary and tertiary structure? This problem has been partially discussed within computer simulations of globular proteins \cite{Dill,Socci}. Some preliminary theoretical results concerning the formation of a  LC-globule can be found in the review \cite{Grosberg}. For a system of fixed composition of stiff parts (helices) and flexible parts, micelle formation has been studied using scaling considerations \cite{Nowak1}. 

In this paper the phase behavior of a helix-coil copolymer is 
discussed by means of a self-consistent field theory (SCFT).
In addition to two- and three-body contact interactions between 
the segments, each segment is also allowed to undergo a 
microscopic transition from flexible to stiff (i.e. being 
part of a helix). To describe this local helix-coil transition 
a grand canonical formalism was used. A formal derivation of 
the grand canonical SCFT free energy functional 
$F[\varphi({\bf r}), \psi({\bf r}, {\bf u}); \mu, \epsilon, \sigma]$ 
is given in \cite{Nowak2}.
We introduced a field 
$\varphi({\bf r})$ associated with a flexible segment at position 
${\bf r}$ and a field $\psi({\bf r}, {\bf u})$ associated with a 
helical (rod-like) segment at position ${\bf r}$ with 
orientation ${\bf u}$. The local densities of flexible and helical segments 
are given by $\rho_{\rm c}({\bf r}) = 1/2\, \varphi({\bf r})^2$ and 
$\rho_{\rm h}({\bf r}) = 1/2 \int d^2 u\, \psi({\bf r}, {\bf u})^2$ respectively.
It is very difficult to deal with the full orientation dependence of $\psi({\bf r}, {\bf u})$.
However, the helices are modeled as stiff rods without chirality, 
so we expect the solution for $\varphi({\bf r})$ and 
$\psi({\bf r}, {\bf u})$ to have an uniaxial symmetry. 
Therefore we choose, without loss of generality, the $z$-axis as the preferred orientation 
and expand the orientation dependence in terms of Legendre polynomials. To lowest 
non-trivial order this expansion reads 
\begin{eqnarray}
\psi({\bf r}, {\bf u}) \approx \left(\frac{1}{4 \pi}\right)^{1/2}
\psi_{0}({\bf r}) + \left(\frac{5}{4 \pi}\right)^{1/2} \psi_{2}({\bf r}) 
P_{2} ({\bf u} \cdot {\bf n}_z).
\label{Uniaxial}
\end{eqnarray}
Using this expansion ${\bf u}$ can be integrated out and the SCFT free energy functional of our 
model is then given by the following form 
\begin{widetext}
\begin{eqnarray}
F[\varphi({\bf r}), \psi_{0}({\bf r}), \psi_{2}({\bf r}); \mu, \epsilon, \sigma] = 
\frac{\mu -\epsilon}{2}\int d^3 r \left[\psi^2_{0}({\bf r}) + \psi^2_{2}({\bf r}) \right]
-\frac{1}{210\,\left(\mu  -\epsilon \right)} \int d^3 r \Bigl\{35\,\psi_{0}({\bf r}) \nabla_r^2 \psi_{0}({\bf r})\nonumber\\
+ 14\,{\sqrt{5}}\,\psi_{0}({\bf r})\Bigl[2\,\partial_z^2 - \partial_x^2 - \partial_y^2 \Bigr]\psi_{2}({\bf r})
+ \psi_{2}({\bf r})\Bigl[25\,\partial_x^2 + 25\,\partial_y^2 + 55\,\partial_z^2\Bigr]\psi_{2}({\bf r})\Bigr\}\nonumber\\
+ \frac{1}{2}\int   d^3r \: \varphi({\bf r})\Bigl[ \mu -  \frac{a^2}{6} \nabla_{r}^2 \Bigr] \varphi({\bf r})
+ \frac{\chi}{4}\int d^3 r \left[ \psi^2_{0}({\bf r}) + \psi^2_{2}({\bf r}) \right]^2
+ \frac{v}{8}\int d^3 r \left[ \varphi^2({\bf r})  + \psi^2_{0}({\bf r}) + \psi^2_{2}({\bf r}) \right]^2\nonumber\\ 
+ \frac{w}{48}\int d^3 r\, \left[ \varphi^2({\bf r})  + \psi^2_{0}({\bf r}) + \psi^2_{2}({\bf r}) \right]^3
- 2\,{\sqrt{\pi \sigma }}\int d^3 r\, \varphi({\bf r})\, \psi_{0}({\bf r}).
\label{Functional}
\end{eqnarray}
\end{widetext}
The numerical prefactors in Eq.(\ref{Functional}) are due to the expansion of the orientation dependence - see Eq.(\ref{Uniaxial}) - 
and the subsequent integration over ${\bf u}$.
%For a detailed exposition  of the model we refer 
%to our recent publication \cite{Nowak} where the grand canonical SCFT free energy  functional $ F[\psi({\bf r}),\psi({\bf r}, {\bf u}); \mu, \epsilon, \sigma]$ (see Eq.(58) in \cite{Nowak}) has been derived. 
The chemical potential $\mu$ 
is used to control the total number of segments of the polymer $N$.  
To model the energy gain due to the formation of hydrogen bonds in helices 
we introduced an energy gain per helical segment $\epsilon$. The cooperativity effect 
in the formation of helices is taken into account by the cooperativity parameter 
$\sigma$. It can be regarded as a fugacity of the interfaces (or junction points) between a helix and 
a flexible part. $\sigma=1$ means no cooperativity effect and $\sigma=0$ 
reflects total cooperativity (i.e. the system can only form a fully flexible chain 
or one long helix). 
The interaction parameters $v$ and $w$ are global two- and three-body interaction constants between all segments, 
whilst $\chi$ controls the 
strength of a selective two-body interaction (due to hydrophobicity) between the helical segments only. 
Further technical details can be found in \cite{Nowak2}.

At this point it is important to emphasize that all interactions in this model are 
point contact interactions. We do not take into account any angle dependent interactions 
between the helical segments which explicitly favor alignment of two helices. 
The total number of segments 
$N$ is given by the sum of all segments in the helical and coil state: $N=N_{\rm h} + N_{\rm c}$, where
\begin{eqnarray}
N_{\rm h} &=& \frac{1}{2} \int d^3r  \:\left[\psi^2_{0}({\bf r}) + \psi^2_{2}({\bf r})\right]\nonumber\\
N_{\rm c} &=& \frac{1}{2} \int d^3 r \varphi^2({\bf r}).
\label{Nh-c}
\end{eqnarray}
Minimization of the functional in Eq.(\ref{Functional}) with respect to 
$\varphi$, $\psi_{0}$ and $\psi_{2}$ yields a set of three coupled differential equations 
for the three self-consistent fields. These differential equations are highly non-linear 
and can only be solved numerically. This was done using the finite element toolkit 
Gascoigne \cite{gascoigne}.

\begin{figure}
\includegraphics[scale=0.5]{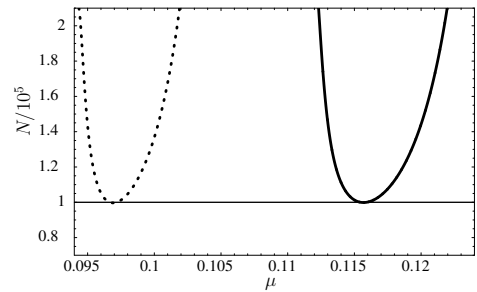}
\caption{$N$ as function of $\mu$ for $w=1.0$, $v=0.2$ and $\sigma=10^{-4}$. The dotted curve corresponds 
to $\epsilon=0.08$ and $\chi=0$. The continuous curve corresponds to $\epsilon=0.1$ and $\chi=0.0138$.}
\label{Nmu}
\end{figure}

The system shows a collapse transition from an open chain to a dense globule. 
In this letter we always consider finite systems, therefore this transition is a crossover 
transition with a finite broadness. Hence we have to define a point during the crossover 
as the transition point. Before we explain how we do this for the coil-globule transition it 
is necessary to clarify  how we generally deal with the chemical potential. In the grand
canonical ensemble the number of particles, here the total number of segments of the 
polymer $N$, is not fixed but its mean value is determined by equilibrium conditions. 
However, in a real experiment  the helix-coil copolymer has a fixed length. In order to ensure this 
fixed length $N$ we tune the chemical potential $\mu$ for each set of physical parameters 
($v,w,\chi,\epsilon,\sigma$) such that the equilibrium value of $N$, calculated by performing 
the integrations in Eq.(\ref{Nh-c}) is equal to the desired one. 
For a given set of parameters 
$N(\mu)$ can be computed and a typical example of this is shown  in Fig.\ref{Nmu}. 
For $\mu \rightarrow 0$ the total number of segments $N$ diverges. This corresponds to the 
$N \sim \mu^{-1}$ behavior of a $\Theta$-solvent chain. The divergence of $N$ at a specific 
value $\mu$ on the right hand side of the minimum corresponds to a fully collapsed infinite globule. 
The minimum of $N (\nu) $ is naturally associated with  the transition point. 
We always fix $N$ for a given set of parameters by tuning $\mu$, so from a plot like the one 
shown in Fig.\ref{Nmu}, we can distinguish whether the system is left of the transition 
point (i.e. in the open chain regime) or right of the transition point (i.e. in the globular 
regime). The transition point is now defined as the set of parameters at which the chosen 
fixed $N$ is equal to the minimum of $N(\mu)$. Both curves in Fig.(\ref{Nmu}) have their minimum at 
$N=10^5$, the value which we have chosen for the results presented below. 
The minimum of the dotted curve corresponds to the first triangle ($\epsilon=0.08$, $\chi=0$) 
in the phase diagram Fig.(\ref{phasediagram}) and the minimum of the continuous curve to the second triangle 
($\epsilon=0.1$, $\chi=0.0138$). 
It is pertinent to note that the characteristic dependence of $N$ on $\mu$ for a simple homopolymer globule 
has first been discussed by Kholodenko and Freed \cite{Kholod}.
For all the results presented  below, 
the global three-body interaction constant is set to $w=1.0$, the global two-body interaction 
constant to $v=-0.2$ and the cooperativity parameter to $\sigma=10^{-4}$.
The SCFT formalism is a reasonable approximation only for systems in a globular state or close to a globular 
state. Therefore $v$ has to be negative in order to ensure that this is always the case. 
Although we do not use the SCFT treatment of our model to describe an open chain of the helix-coil copolymer, 
it is perfectly possible with the definition given above to calculate the transition line between 
the globular state and the open chain state.  
 
The really astonishing feature of this model is that it also shows a crossover transition from a 
disordered amorphous globule with low or mid fraction of helical segments 
$\Theta_{\rm h}=N_{\rm h}/N$ 
to an ordered liquid-crystalline globule with very high fraction of helical segments. 
This transition occurs despite the fact that there is no explicit angle dependent alignment interactions. 
The transition is triggered by a subtle interplay of the entropy contribution (surface energy), represented by 
the derivative terms in Eq.(\ref{Functional}), and bulk interaction energy, represented by the 
$\chi$-term. It is well known \cite{Grosberg1} that in a simple homopolymer globule the surface energy has an entropic nature (since the conformational set of surface segments is constrained) and is isotropic. As one can see from Eq. (\ref{Functional}) in our case the surface energy is {\it anisotropic}, so that after a proper inspection \cite{Subseq} one can ensure that the surface tension in the $xy$-direction is smaller than the one in  the  $z$-direction. That is why the system tries to maximize its lateral  surface in $xy$-directions and  minimize it in $z$-direction, i.e. a nematic, cigar shaped, LC-globule occurs.

\begin{figure}
\includegraphics[scale=0.63]{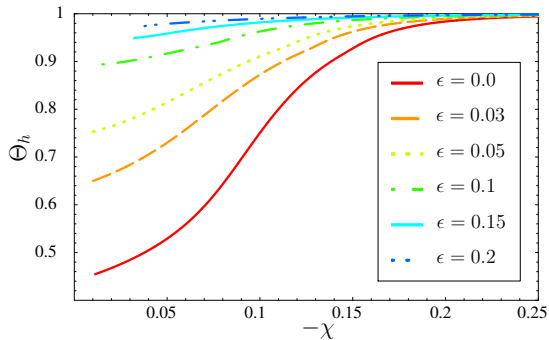}
\caption{Fraction of helical segments $\Theta_h$ as a function of strength $-\chi$ of the attractive two-body interaction 
between the helical segments only for different values of the energy gain per helical segment $\epsilon$.} 
\label{fracH}
\end{figure}

\begin{figure}
\includegraphics[scale=0.66]{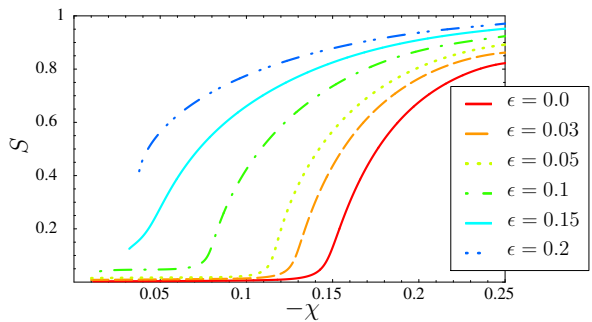}
\caption{Nematic order parameter $S$ as a function of $-\chi$ for different values of $\epsilon$.}
%\label{nem-fracH}
\label{nem}
\end{figure}

To measure orientational order in the system we define the nematic order parameter $S$ as 
follows (see e.g. \cite{Prost})
\begin{eqnarray}
S &\equiv&  \frac{1}{3 N}\int d^3r \int d^2 u \: P_2 
(\cos \theta) \psi^2({\bf r}, {\bf u}) \\\nonumber
&=& \frac{1}{N}\int d^3r \left(\frac{2}{\sqrt{5}} \:  \psi_{2}({\bf r})
\left[\psi_{0}({\bf r}) + \sqrt{5} \psi_{2}({\bf r})\right]\right).
\label{D2}
\end{eqnarray}
Fig.(\ref{fracH}) shows the increase in fraction of helical segments $\Theta_{\rm h}$ with 
$|\chi|$ during the transition and Fig.(\ref{nem}) shows the simultaneous onset of a finite nematic order parameter $S$. 
The onset of the transition is shifted to lower values of $|\chi|$ with increasing energy gain 
per helical segments $\epsilon$. This is due to an increase of bulk interaction energy for fixed $\chi$ 
with increasing number of helical segments. The curves for $\epsilon=0.2$ and $\epsilon=0.15$ in Figs.(\ref{fracH},\ref{nem}) 
start at non-zero values of $\chi$. These values of $\chi$ correspond to the transition point between open chain state 
and globular state, i.e. the triangles in Fig.(\ref{phasediagram}). 
The transition point for the crossover transition from an amorphous to a nematic LC-globule has to be defined in a reasonable way. 
We define these transition points as the inflection points of the $S(\chi)$-curves in Fig.(\ref{nem}).   

\begin{figure}
\includegraphics[scale=0.15]{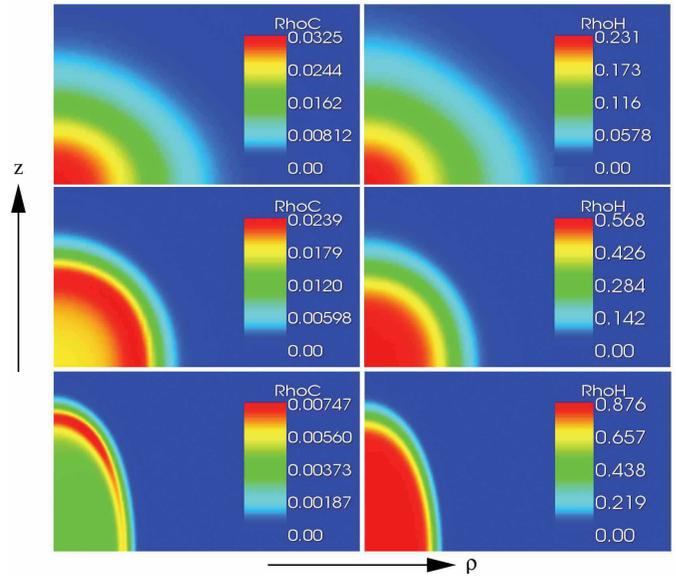}
\caption{The density of the flexible segments is shown on the left and the density of the helical 
segments on the right. $\epsilon=0.1$ for all plots. $\chi=-0.0138$ in the top line, $\chi=-0.0812$ 
in the middle line and $\chi=-0.18$ in the bottom line.}
\label{dens}
\end{figure}

\begin{figure}
\includegraphics[scale=0.7]{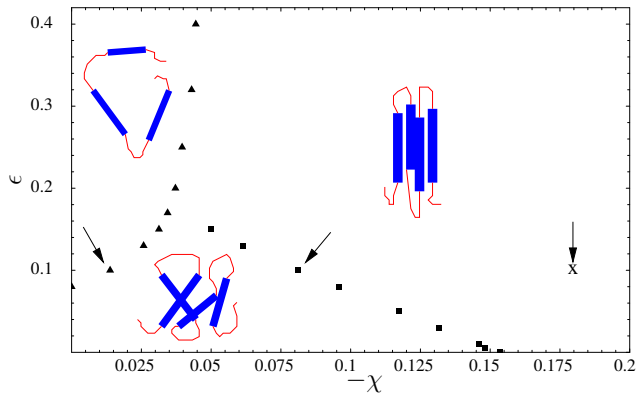}
\caption{Phase diagram of a helix-coil copolymer. The upper left area corresponds to an open chain, 
the lower left area to an amorphous globule and the right area to a nematic LC-globule.
The little arrow to the left indicates the point in the phase diagram which corresponds to the top two 
pictures in Fig.(\ref{dens}). The arrow in the middle corresponds to the middle 
two pictures and the arrow to the right to the bottom ones.}
\label{phasediagram}
\end{figure}

To demonstrate how the shape of the globule changes during the transition we show in Fig.(\ref{dens}) 
a color-coded plot of the local density in $\varrho$-$z$ space, 
where $\varrho=\sqrt{x^2+y^2}$. The center of the globule is in the bottom left corner. $\varrho$ is 
increasing from left to right and $z$ is increasing from bottom to top.
Red means high density and dark blue zero density. 
The top two pictures show the density profile at the transition point between open chain and amorphous globule.
At this point the system is spherical and has a very broad surface layer of decaying density resembling the characteristics of 
a random walk. Although the density of the helical segments shown on the right is higher than the density 
of the flexible segments shown on the left, their distribution and the shape of the profile is very similar.  
The middle two pictures show the density profile at the transition point between amorphous globule and LC-globule. 
The system adopts a slightly cylindrical shape indicating the onset of nematic order. It can also be seen that the 
density maximum of the flexible segments is not in the center of the globule denoting a repulsion of flexible 
segments from the center to the surface layer. The surface layer is now much narrower. 
The bottom two pictures show the density profile deep in the 
nematic LC-globule phase. The globule has developed a strongly asymmetric cylindrical shape indicating strong 
nematic order. The repulsion of flexible segments from the center towards the surface layer can be 
seen clearly and the surface layer is now very narrow.

The results shown in Fig.(\ref{nem}) allow us to compute a complete phase diagram of 
a helix-coil copolymer in $\epsilon-\chi$ space, see Fig.(\ref{phasediagram}).
The triangles are the transition points between open chain and globule. The squares are the transition points 
between amorphous globule and LC-globule. 
Note, the points plotted in the phase diagram, Fig.(\ref{phasediagram}), are what we defined 
above as the transition points of rather broad crossover transitions. Therefore the boundaries in the 
phase diagram have to be understood as ''center lines'' of broader regions in which the crossover from 
one phase to the other occurs. Although the qualitative shape of the phase diagram stays the same, the 
position of the transition lines changes with $N$, $\sigma$ and $v$. If $|v|$ is increased the open chain 
region of the phase diagram becomes smaller (and disappears eventually). With decreasing $\sigma$ the 
transition from a disordered to an ordered globule becomes sharper and the transition line between 
these two regions is shifted to smaller values of $|\chi|$. Since this transition occurs due to an interplay 
between surface energy and bulk interaction energy, the transition also becomes sharper for decreasing 
system size $N$, which, at first sight, is a rather unusual and surprising behavior. For smaller systems 
the surface energy plays a bigger role and therefore leads to a sharper transition. 
For $N \rightarrow \infty$ the ordered globule phase finally disappears, since the surface contributions to 
the free energy vanish for infinite systems. A detailed discussion of the modification of the phase diagram 
will be given in a subsequent publication \cite{Subseq}

In summary, we presented a SCFT for a multiblock helix-coil copolymer chain based on grand canonical ensemble considerations.  
The minimization of the corresponding free energy functional - Eq.(\ref{Functional}) - has been done numerically and it has been shown that three phase states can be clearly seen: open helix-coil chain, amorphous globule and  nematic LC-globule. It is a novel result that the formation of a LC-globule occurs without explicit alignment interactions between the helical parts. It is the entropical surface tension anisotropy which drives the globule in the nematic LC-state in order to maximize the density of the hydrophobic helical segments. In the presence of an explicit attractive alignment interaction (which can be taken into account by a Maier-Saupe term) one only sees an enhancement of this effect \cite{Subseq}. We believe that this model provides a relatively simple example of the interplay between secondary and tertiary structure in homopolypeptides. It can also give a simple explanation for the formation of helix bundles in certain globular proteins. Both, simulations \cite{Zhou} and experiments \cite{Ptitsyn} show that proteins can adopt not only the native state and completely denaturated state (open chain) but also so-called  premolten and molten globular states. For helix-bundle proteins the premolten globule, which does not show any order of the helices, corresponds to our amorphous globule. The molten globule with ordered helices but without native contacts corresponds to our LC-globule.   
The chain length of proteins is typically of the order of $N \sim 10^2-10^3$ instead of $N \sim 10^5$. However, as mentioned above the crossover from an amorphous globule to a nematic LC-globule with aligned helices becomes even sharper for shorter chains \cite{Subseq}. This indicates that our model indeed provides a simple explanation for the formation of helix bundles.  

C. Nowak would like to thank T. Richter for invaluable help on how to use 
Gascoigne and D. Andrienko for fruitful discussions. V.G. Rostiashvili and T.A. Vilgis appreciate the financial support of
the German Science Foundation (SFB 625). 

%%%%%%%%%%%%%%%%%%%%%%%%%%%%%%%%%%%%%%%%%%%%%%%%%%%%%%%%%%%%%%%%%%%%%%%%%%
%%%%%%%%%%%%%%%%%%%%%%%%%%%%%%%%%%%%%%%%%%%%%%%%%%%%%%%%%%%%%%%%%%%%%%%%%%%

%%%%%%%%%%%%%%%%%%%%%%%%%%%%%%%%%%%%%%%%%%%%%%%%%%%%%%%%%%%%%%%%%%%%%%%%%%%

\end{document}